\documentclass[%
 aip,groupedaddress,
 jmp,%
 amsmath,amssymb,
preprint,
]{revtex4-1}

\usepackage{graphicx}% Include figure files
\usepackage{float}
\usepackage{dcolumn}% Align table columns on decimal point
\usepackage{bm}% bold math
\usepackage{amsmath,amsfonts,amsthm,amssymb}
\usepackage{xcolor}
\def\apjl{{\em The Astrophsical Journal Letters}}

\def\grl{{\em Geophys. Res. Lett.}}
\def\jgr{{\em J. Geophys. Res.}}
\def\prl{{\em Phys. Rev. Lett.}}

\def\nat{{\em Nature}}

\def\sci{{\em Science}}
\def\pp{{\em Plasma Physics}}
\def\pop{{\em Phys. Plasma}}
\def\pof{{\em Phys. Fluid}}

\begin{document}

\preprint{AIP/123-QED}
\title{Two-Fluid Description of Wave-Particle Interactions in Strong Buneman Turbulence }% Force line breaks with \\
%\thanks{Footnote to title of article.}

\author{H. Che}
%\author{B. Author}%
% \email{Second.Author@institution.edu.}
\affiliation{NASA/Goddard Space Flight Center, Greenbelt, MD, 20771, USA}
%Authors' institution and/or address%\\This line break forced with \textbackslash\textbackslash
% \homepage{http://www.Second.institution.edu/~Charlie.Author.}
%\affiliation{%
%Second institution and/or address%\\This line break forced% with \\
%}%


\begin{abstract}
To understand the nature of anomalous resistivity in magnetic reconnection, we investigate turbulence-induced momentum transport and energy dissipation while a plasma is unstable to the Buneman instability in force-free current sheets. Using 3D particle-in-cell simulations, we find that the macroscopic effects generated by wave-particle interactions in Buneman instability can be approximately described by a set of electron fluid equations. We show that both energy dissipation and momentum transport along electric current in the current layer are locally quasi-static, but globally dynamic and irreversible. Turbulent drag dissipates both the streaming energy of the current sheet and the associated magnetic energy. The net loss of streaming energy is converted into the electron component heat conduction parallel to the magnetic field and increases the electron Boltzmann entropy. The growth of self-sustained Buneman waves satisfies a Bernoulli-like equation that relates  the turbulence-induced convective momentum transport and thermal momentum transport. Electron trapping and de-trapping drive local momentum transports, while phase mixing converts convective momentum into thermal momentum. The drag acts like a micro-macro link in the anomalous heating processes. The decrease of magnetic field maintains an inductive electric field that re-accelerates electrons, but most of the magnetic energy is dissipated and converted into the component heat of electrons perpendicular to the magnetic field.  This heating process is decoupled from the heating of Buneman instability in the current sheets. Ion heating is weak but ions plays an important role in assisting energy exchanges between waves and electrons. Cold ion fluid equations together with our electron fluid equations form a complete set of equations  that describes the occurrence, growth, saturation and decay of the Buneman instability. 
\end{abstract}

\maketitle

\section{Introduction}
\label{intro}
Magnetic reconnection is a process in plasma where magnetic field topology rearranges and magnetic energy is converted into the  energy of plasma. A current layer at the contact surface of oppositely directed magnetic fields is a standard configuration of magnetic reconnection. 
%(Fig.~\ref{illu}).
 Such magnetic field configuration and the associated current layers have been observed in the magnetopause and magnetotail of the Earth\cite{gosling96jgr,russell97grl,phan07prl,ange08sci,pu10jgra}, in the corona of the Sun, \cite{wang07apjl,liu10apjl,gosling13apjl} and should be common in astrophysical environments.

For magnetic reconnection to occur, the ideal magnetohydrodynamics (MHD) frozen-in condition $\mathbf{E}+\mathbf{U}\times \mathbf{B}=0$ must be broken. This takes place in the so-called diffusion regions where the ions and electrons demagnetize. The dimension of the electron (ion) diffusion region is of order $d_e = c/\omega_{pe}$ ($d_i = c/\omega_{pi}$), where $\omega_{pe}$ ($\omega_{pi}$) is the plasma electron (ion) frequency. Single fluid MHD equations are obtained from two-fluid equations under the assumption of low wave frequency ($<<\Omega_i$) and high collision rate. In the diffusion regions, the frequency of plasma waves range from $\sim\Omega_i$ to $\Omega_e$. Thus single fluid MHD equations are generally not valid in diffusion regions, and two-fluid equations are required to describe the macroscopic processes in the diffusion regions. The two-fluid equation for particle species $s$ ($s$ is either electron or ion) is: 
\begin{equation}
q_s n_s\mathbf{E}  + \dfrac{1}{c}\mathbf{j}_s \times \mathbf{B} =\partial_t \mathbf{p}_s+ \nabla\cdot (\mathbf{p}_s \mathbf{U}_s)
 +\nabla \cdot \mathbb{P}_s +\eta \mathbf{j}, 
\label{2flu}
\end{equation}
where $\mathbf{j} \equiv \mathbf{j}_e+\mathbf{j}_i$, $\mathbf{p_s}\equiv m_s n_s \mathbf{U_s}=\mathbf{j_s}/q_s$, $\eta$ is the collisional resistivity,  $\mathbb{P}_s$ is the pressure tensor, $q=e$ for ions and $q=-e$ for electrons. The merging of magnetic field lines will not occur until both the ion and electron frozen-in conditions are broken, i.e. $\mathbf{E}  + \mathbf{U}_s \times \mathbf{B}/c \neq 0$. 

Turbulence is often observed to associate with magnetic reconnections in magnetosphere, solar flare and lab magnetic reconnection experiments\cite{gosling96jgr,russell97grl,bale02grl,matsumoto03grl,cattell05jgr,wang07apjl,phan07prl,ange08sci,eastwood09prl,pu10jgra,mozer11pop,
che11nat,fox12pop,liu10apjl,gosling13apjl,wendel13jgr}. In diffusion regions, kinetic turbulence is common. Turbulence-induced heating, commonly called  ``anomalous resistivity'', is a widely invoked mechanism to facilitate fast magnetic reconnection \cite{sagdeev62book,huba77grl,kulsrud05pop}. However, what role anomalous resistivity plays in magnetic reconnection is still not fully understood and is a question of great interest\cite{drake03sci,cattell05jgr,che10grl,mozer11jgr,prit13pop}. Kinetic turbulence causes various macroscopic processes. Understanding these processes is key to find out the influence of kinetic turbulence on reconnection. The essential process in kinetic turbulence  is wave-particle interactions, but the effects of wave-particle interactions are not included in the fluid equations. Understanding the macroscopic effects caused by wave-particle interactions and incorporating them into fluid equations is the goal of this study. 
 
Primarily two types of approaches exist in incorporating kinetic effects into fluid equations.  The simplest method is  {\it parametrization}. Anomalous resistivity is written as an effective resistivity $\eta_{eff}$ and the resistive term in Eq.~(\ref{2flu}) becomes $\eta_{eff} \mathbf{j}$. This parametrization does not distinguish the underlying physics between anomalous resistivity and collisional resistivity. The second approach considers the influence of weak kinetic effects on ion scale where ion finite Larmor radius corrections and Landau-damping effects for low frequency waves are important\cite{lee86pof,hamm90prl,passot04pop,brizard08pop}. This method cannot be applied to strong kinetic turbulence, and it ignores wave-electron interactions. The electron dynamics is not negligible on both ion and electron scales, in particular in electron diffusion region of reconnection where magnetic field lines break. In this paper, we approach this problem with a novel method using particle-in-cell (PIC) simulations. We will focus on strong Buneman turbulence and electron dynamics.
 
Buneman instability is common in magnetic reconnection, driven by electron streams around x-lines .\cite{drake03sci, che10grl,khot10prl,che11nat}  It is an electrostatic instability that occurs when the relative drift between ions and electrons is larger than the electron thermal velocity.\cite{buneman58prl} In our earlier paper (Che et al. 2013, Paper I hereafter)\cite{che13pop}, we used PIC simulations to investigate the mechanism of fast electron heating in strong Buneman instability. We found  that the fast energy exchange between waves and electrons is achieved by the adiabatic motion of trapped electrons. The energy gained from waves by these trapped electrons is converted into heat through trapping and de-trapping processes. In this paper, we use the same PIC simulation to investigate the macroscopic effects caused by  strong Buneman instability. We show that, besides anomalous heating, macroscopic momentum transports are also induced.  It is found that a Bernoulli-like equation governs the energy exchange between waves and the electrons, and links microscopic wave-electron momentum exchange to macroscopic momentum transports. This localized quasi-static equation couples with the equation of anomalous heating (which is a global effect) to form a set of fluid equations that describe Buneman instability. More interestingly, the associated magnetic energy is dissipated through the heating of electrons in directions perpendicular to the magnetic field. This process is decoupled from the dissipation of the kinetic energy of the electron stream. While turbulence-induced friction or drag is shown to play a similar role in turbulence heating as collisions do in joule heating, we found that the heating rate of Buneman turbulence depends on the changing rate of the kinetic energy density rather than on the kinetic energy density as in joule heating. Another new finding is that strong Buneman turbulence naturally truncates the electron momentum equation and provides the closure for pressure. Ions play an important role in assisting the energy exchange between waves and electrons even though ions are weakly heated. The role of ions in Buenman instability can be simply described by cold ion fluid. The ion equations together with those of electrons form a complete kinetic description of strong Buneman instability.
 
%{\it Drag} is the source of the various macroscopic effects in Buneman instability, hence we first discuss how to incorporate drag into Eq.~\ref{2flu}. We then use the spatial characteristics of turbulence as well as simulations to investigate the global and localized macroscopic effects. 

\section{Incorporating Turbulence Drag into Two-fluid Equations }
\label{drag2flu}
Electrostatic instabilities satisfying $\mathbf{k}\times \mathbf{B}=0$ and $\delta \mathbf{B}=0$ produce self-sustained electric field $\delta \mathbf{E}$ through trapping of charged particles, i.e. $\nabla \cdot \delta \mathbf{E}=\delta n_e +\delta n_i$. Turbulence-induced friction is produced by local interactions between trapped particles and the self-sustained electric field, i.e. $q\delta n_s\delta \mathbf{E}$, known as electron or ion {\it drag}. Drag is the only force induced in an electrostatic instability and is the source of all  macroscopic effects.  
In this section, we incorporate drag into fluid equations so that Eq.~\ref{2flu} includes the kinetic electrostatic turbulence friction. 
 
Instability-driven turbulence is characterized by fast and slow varying fluctuations on different spatial scales. Thus it is useful to split each physical quantity $A$ into a fast turbulent fluctuation $\delta A$ and a mean value over some large region with dimension $L>> 1/k_{p}$ (where $k_{p}$ is the wave number of fastest-growing mode of the instability) in which the underlying physical conditions are similar:  
\begin{gather}
\begin{split}
 A  =  \langle A \rangle + \delta A,  \\
 \langle \delta A \rangle=0.
 \end{split}
 \label{split}
\end{gather}
In the case of one dimensional turbulence, the spatial average is defined as  
\begin{equation}
\langle A \rangle  \equiv   \dfrac{\int_{-L/2}^{L/2} w(x^{\prime})  A(x- x^{\prime})dx^{\prime}}{\int_{-L/2}^{L/2} w(x^{\prime}) dx^{\prime}} 
\label{ta}
\end{equation}
and $w(x^{\prime})$ is the weighting function. 

 We assume the background electric field $\mathbf{E}_0=0$. Since drag is only related to fluctuations of density and electric field, we split $n_s=\langle n_s\rangle + \delta n_s$ and $\mathbf{E}=\langle \mathbf{E}\rangle +\delta \mathbf{E}$. Using the facts that $|\delta n_s | / \langle n_s \rangle \lesssim 1$ and $|\langle \mathbf{E}\rangle|/| \mathbf{E}|<<1$ for strong electrostatic turbulence, we have $n_s \mathbf{E}= \delta n_s \delta \mathbf{E}+\langle n_s\rangle \mathbf{E}(1 + \delta n_s\langle \mathbf{E}\rangle)/(\langle n_s\rangle \mathbf{E})\approx \delta n_s\delta \mathbf{E} + \langle n_s\rangle \mathbf{E}$. Inserting these into Eq.~(\ref{2flu}) we obtain:
\begin{equation}
%\begin{multline}
  \mathbf{E} + \mathbf{U}_s \times \mathbf{B}/c =  \mathbf{D}_{s} +\dfrac{m_s}{q_s}(\partial_t  \mathbf{U}_{s}  + \mathbf{U}_{s} \nabla \cdot \mathbf{U}_{s}) +\dfrac{1}{q_s\langle n_s\rangle} \nabla \cdot \mathbb{P}_{s},
\label{turbflu}
%\end{multline}
\end{equation}
where $\mathbf{D}_s \equiv -\delta n_s \delta \mathbf{E}/\langle n_s\rangle$ is the drag. If there is no turbulence, then $\mathbf{D}_s \approx 0$ and the equation reduces to Eq. (\ref{2flu}). It is worth noting that drag D is local and the mean bracket $\langle\rangle$ does not appear. We used the approximation $ n_s/\langle n_s\rangle(\partial_t  \mathbf{U}_{s}  + \mathbf{U}_{s} \nabla \cdot \mathbf{U}_{s})\approx \partial_t  \mathbf{U}_{s}  + \mathbf{U}_{s} \nabla \cdot \mathbf{U}_{s}$ in Eq.~(\ref{turbflu}). The reason is that $\delta n_s$ fluctuated around zero and does not have direct correlations with $\partial_t \mathbf{U}_s$ and $\nabla \cdot\mathbf{U}_s$, thus its contribution to the inertial terms is negligible. 

Electron dynamics dominate in the diffusion region of magnetic reconnection. The role of ions in Buneman instability on the other hand is to facilitate the exchange of momentum between electrons the waves, and its dynamics is simple.  

Drag is the source of kinetic turbulence macroscopic effects. While Eq.~\ref{turbflu} includes the effects of drag, it is still unknown how to calculate the drag. In the  following sections, we will find an equation to describe the evolution of Buneman waves and an energy equation to provide a closure for the pressure through investigating what momentum transports are produced by drag using PIC simulations.  
\section{Spatial-Averaged Electron Equation for Collisionless Electrostatic Turbulence}
\label{ohm}

To investigate momentum transports and energy transfer, we need to separate ``global" from ``local" effects produced by drag generated by local wave-particle interactions. After spatial averaging some quantities are zero while others are non-zero. We call the effects produced by quantities with non-zero spatial average  \textit{global} effects, and the effects produced by quantities with zero spatial average \textit{local} effects. We consider only collisionless plasma thus $\eta =0$. We perform spatial average on Eq.~(\ref{turbflu}) to investigate the global effects. Taking into account of the fact that the spatial and temporal differential operators commute with the spatial average operation defined by Eq.~(\ref{ta}), we obtain:     
 \begin{widetext}
 \begin{flalign}
\langle\mathbf{E}\rangle=-\dfrac{m_{e}}{e} (\partial_t \langle\mathbf{U}_{e}\rangle+ \langle\mathbf{U}_{e}\cdot \nabla\mathbf{U}_{e}\rangle) 
- \dfrac{1}{c} \langle\mathbf{U}_{e}\rangle  \times \langle\mathbf{B}\rangle 
-\dfrac{\nabla \cdot \langle\mathbb{P}_{e}\rangle}{e\langle n_{e}\rangle}
+\langle\mathbf{D}_e\rangle. 
\label{eturb1}
\end{flalign}
\end{widetext}
This equation governs the global/macroscopic properties of the plasma when Buneman instability is present. The combination of the first two terms on the right-hand side of Eq.~(\ref{eturb1}) is inertia. We call  $m_{e} \partial_t \langle\mathbf{U}_{e}\rangle/e$ acceleration, and $m_{e} \langle\mathbf{U}_{e} \cdot \nabla\mathbf{U}_{e}\rangle/e$ mean convective momentum transport.  The mean drag is $\langle\mathbf{D}_e\rangle  \equiv - \langle \delta n_{e} \delta \mathbf{E} \rangle/\langle n_{e}\rangle$, and the mean anomalous thermal momentum transport $-\nabla \cdot \langle\mathbb{P}_{e}\rangle/(e\langle n_{e}\rangle)$.  $\nabla \cdot \langle\mathbb{P}_{e}\rangle$ includes second order correlations caused by turbulence.  Since we do not introduce approximations that require fast varying terms to be small, Eq.~(\ref{eturb1}) applies to both weak and strong turbulence.

 In the following sections we use our 3D PIC simulation to study each of the terms in Eq.(\ref{turbflu}) and (\ref{eturb1}) in the presence of Bunamen instability to obtain anomalous momentum transports and energy conversion relations with nearly zero ion drift.
 
 \section{ Energy Dissipation and Momentum transports in Buneman Turbulence}
 \label{sec3}
 
 \subsection{ Simulation} 
 \label{simu} 
The 3D PIC simulation we use in this paper has been discussed in detail in Paper I and here we briefly summarize.  The simulation is set-up to mimic the current sheet at the x-line in a guide-field magnetic reconnection when Buneman instability occurs. The coordinate system is chosen so that the current layer lies in the x-z plane.  The mid-plane of the current layer has $y=0$, and the guide magnetic field is in $z$-direction. No external perturbations are applied to initiate magnetic reconnection, and reconnection does not develop spontaneously during the simulation. The initial magnetic field has the form $B_x/B_0=\tanh[(y-L_y/2)/w_0]$, where $B_0$ is the asymptotic amplitude of $B_x$;  $w_0$ and $L_y$ are the half-width of the initial current sheet and the box size in $y$-direction, respectively. The guide magnetic field $B_z^2 = B^2-B_x^2$ is chosen so that $\vert B\vert$ is constant. We choose the following parameters for our simulation:  the mass ratio between ion and electron $m_i/m_e=100$,  $w_0 = 0.1 d_i=d_e$,  $\vert B\vert=\sqrt{26} B_0$,  and the initial isotropic and uniform temperature $T_{e0} =T_{i0} =0.04 m_i c_{A0}^2$, where  $c_{A0}= B_0/(4 \pi n_0 m_i)^{1/2}$ is the asymptotic ion Alfv\'en wave speed. Within the current layer, the electron cyclotron frequency $\Omega_{e}=eB/cm_e\sim 509 \Omega_{i0} \sim 0.636 \omega_{pe}$, where $\Omega_{i0} \equiv eB_0/(m_i\thinspace c)$. The simulation domain has dimensions $L_x \times L_y \times L_z= 1 d_i \times 1 d_i \times 2 d_i$, with periodic boundary conditions in $x$ and $z$, and a conducting boundary condition in $y$. The cell numbers in $x$, $y$ and $z$ directions are $512\times 512\times 1024$. The initial electron drift have velocity $v_{de}  \sim 9 c_{A0} \sim 3 v_{te0}$ ($v_{te0}$ is the electron thermal velocity) along $z$, which is large enough to trigger Buneman instability. The initial ion drift is about 0.9 $v_{A0}$ is only tenth of the electron drift and also much smaller  than $v_{te0}$. Thus in the following we neglect the ion's drfit.
 
 \begin{figure}
\includegraphics[scale=0.8,trim=0 60 0 0,clip]{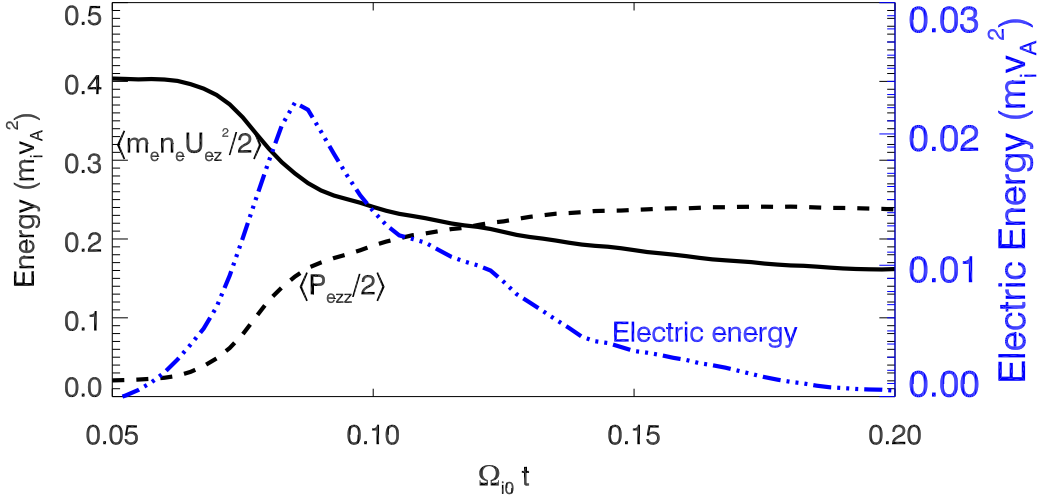}
\caption{Time evolution of $\langle  P_{ezz}/2\rangle$ (black dashed line), the kinetic energy of electron beams $m_e \langle n_e U_{ez}^2/2\rangle$ ( solid black line, where $U_{ez} = - j_{ez}/(e n_e)$), and the electric energy (blue dots-dashed line, scales shown on the right side of the box in blue color).}
\label{diss2}
\end{figure}

Buneman instability starts at $\Omega_{i0} t \sim 0.025$. The growth rate $\gamma\sim 0.12 \omega_{pe}\sim 96 \Omega_{i0}$ in our simulation is close to the Buenman growth rate in cold plasma limit $\sqrt{3}/{2}(m_e/2m_i)^{1/3}\omega_{pe}$.  The instability saturates at $\Omega_{i0} t \sim 0.078$ when the electric field reaches its peak of $40 E_0 -60 E_0$, where $E_0= c_{A0} B_0/c$. The electric field then decays to half of the peak value at $\Omega_{i0} t \sim 0.125$. Around the time when Buneman instability saturates (roughly between $\Omega_{i0} t = 0.075$  and 0.125), the electron temperature exhibits a rapid increase. Since the electron bounce rate $\omega_b=k_0\sqrt{e\phi/m_e}\sim \omega_{pe}$ is much larger than the growth rate $\gamma\sim 0.013\omega_{pe}$ near saturation, the energy exchange between waves and electrons is caused by the nearly adiabatic motion of electrons. The continuous non-adiabatic trapping and de-trapping of electrons with velocities $-v_{de} \lesssim v \lesssim v_{de}$ convert the energy gained from waves into electrons' thermal energy, resulting in a rapid increase of the $zz$ component of electron temperature and a rapid decrease of kinetic energy of electron streams. As shown in Fig.\ref{diss2}, from $\Omega_{i0} t \sim$ 0.075 to 0.1, the kinetic energy density of the electron streams $W_k=m_e\langle n_e U_{ez}^2/2\rangle$ decreases from 0.4 to 0.2 and the component of the electron pressure $P_{ezz}/2$ increases from 0.02 to 0.2 and $ \triangle P_{ezz}\sim \triangle W_k$ (A detailed analysis of the heating mechanism can be found in Paper~I). 
\begin{figure}
\centering
\includegraphics[scale=0.8, trim=0 10 0 60,clip]{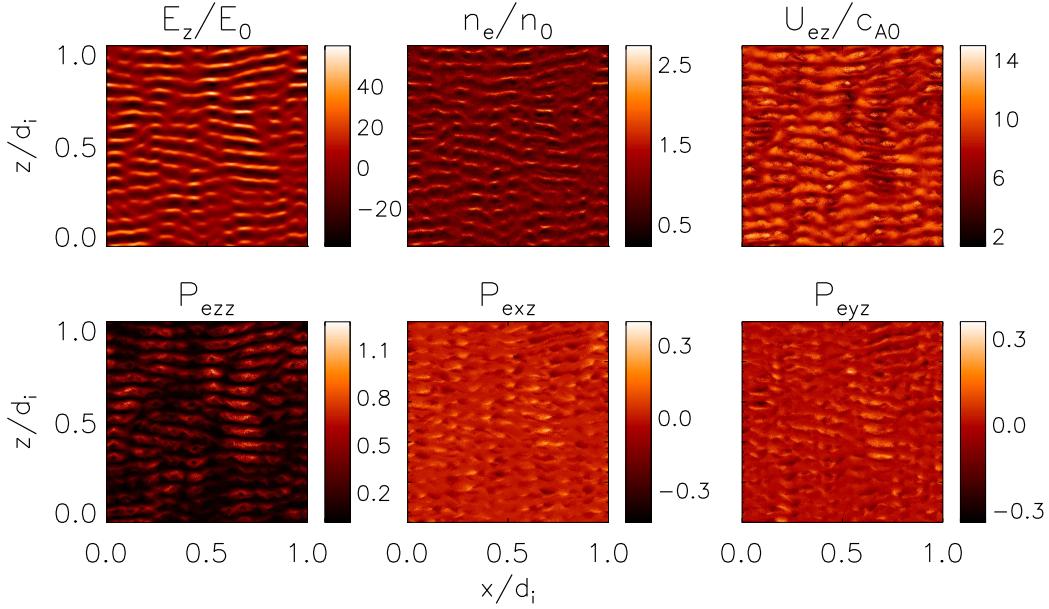} 
\caption{Physical quantities at $\Omega_{i0} t =0.075$ in the mid-plane of the current sheet. }
\label{xz}
\end{figure} 

In Fig.~\ref{xz} we show the electric field $E_z$,  electron density $n_e$, electron fluid velocity $U_{ez}$ and components of pressure in the mid-plane of the current layer at $\Omega_{i0} t \sim 0.075$ when the Buneman instability reaches its peak.  Electrostatic waves $E_z$ propagate along z and form solitary waves. Electron trappings at the locations of intense electric field are strong and electron densities are high. The correlation between  density and electric field causes turbulence drag. Wave patterns of  pressure components and $U_{ez}$ also follow that of the electric field, indicating that the variation of pressure and velocity along $z$ are modulated by the motion of  trapped electrons.  

In Fig.~\ref{xz} it is obvious that the coherent localized electric fields parallel to $z$ form uniformly in the mid-plane of the current layer with no preferred locations. The length of wave pattens along $z$  is close to the wavelength of the fastest Buneman mode $\sim 2\pi  v_{de}/\omega_{pe}\sim 0.08 d_i$. This length is much smaller than the simulation box size $L_z=2 d_i$. We thus can apply spatial average along $z$ over the simulation box to investigate the spatial averaged Ohm's law.

We also use average over x-direction. This is because Buneman waves are parallel to $z$, and the translational symmetry in $x$ direction of the initial set-up guarantees the Buneman waves along $x$-direction are independent realizations of the same physical process. Small variations are found in the solitary waves in Fig.~\ref{xz} that break the alignment of wave patterns in $x$-direction. But it should be noticed that $x$-average is conceptually different from the $z$-average we have applied. We employ $x$-average as a method to reduce noise in the simulation. In the following all quantities are $x$-averaged if not explicitly pointed out (our results are essentially the same without applying $x$-average).

\subsection{Global non-static Effects: Drag Force, Mean Electric Field and the Deceleration of Electron Stream}
\label{sxy}
In this section, we use our simulation to study the $z$-averaged Ohm's law in the thin current layer. We apply average over $[0, L_z]$ thanks to the strong guide field in $z$-direction. If the guide field is weak, the spatial average should be performed along more oblique magnetic field lines since the electrostatic instability is parallel to the magnetic field. We focus on the $z$-component of Eq. (\ref{eturb1}) since Buneman instability grows nearly parallel to $z$ and the most important physics can be learnt by studying the $z$-component of the spatial averaged Ohm's law:
%\begin{widetext}
%\begin{flalign}
\begin{equation}
\langle E_z\rangle=-\dfrac{m_e}{e}(\partial_t \langle U_{ez}\rangle+ \langle \mathbf{U} \cdot \nabla U_{ez}\rangle) -\dfrac{1}{c}(\langle \mathbf{U}_{e\perp}\rangle \times \langle\mathbf{B}_{\perp}\rangle )_z -\dfrac{\nabla\cdot \langle\mathbb{P}_{e\perp z}\rangle}{e\langle n_e\rangle}+\langle D_{ez}\rangle.
\label{eturbz}
\end{equation}
%\end{flalign}
%\end{widetext}

 The terms in Eq. (\ref{eturbz}) related to pressure  $\mathbb{P}_{e\perp z}$  are simplified to $\nabla\cdot \mathbb{P}_{e\perp z}=\partial_x \mathbb{P}_{exz}+\partial_y \mathbb{P}_{eyz}$.
 \begin{figure}
\includegraphics[scale=1, trim=120 40 0 0,clip]{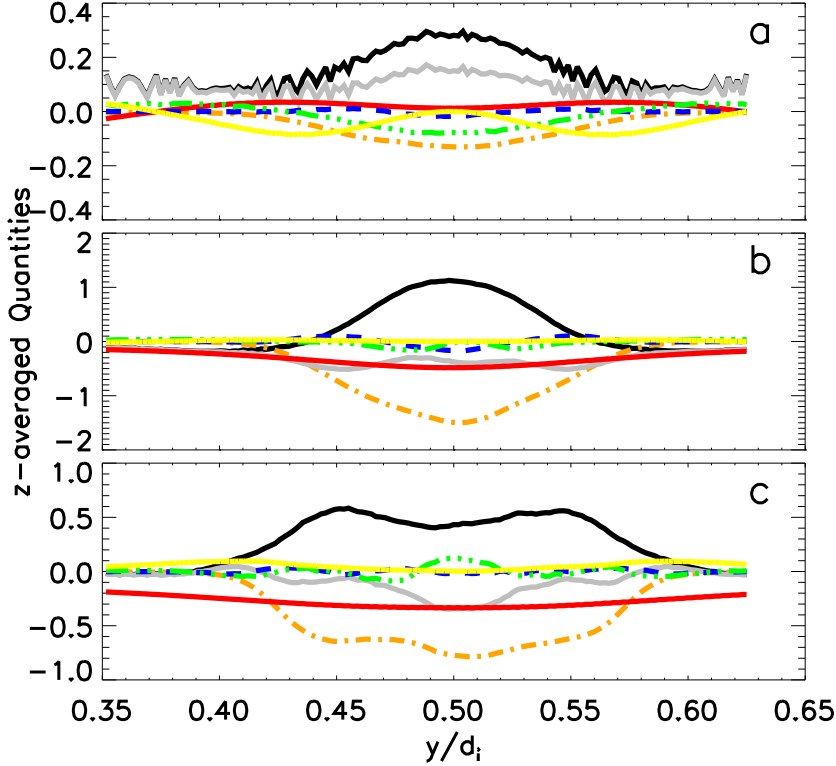} 
\caption{Each of the terms in Eq.(\ref{eturbz}) as a function of y at $\Omega_{i0} t=$ 0.05 (panel a), 0.075 (panel b), and 0.1 (panel c). Red solid lines: the mean electric field $\langle E_z \rangle$;  Black solid lines: inertia $-m_e \partial_t \langle U_{ez}\rangle/e$;  Orange dot-dashed lines: drag $\langle D_{ez}\rangle$; Gray solid lines: $-m_e\partial_t \langle U_{ez}\rangle/e +\langle D_{ez}\rangle$; Green three-dots-dashed lines: the divergence of non-diagonal pressure $-\nabla\cdot \langle \mathbb{P}_{e\perp z}\rangle/(e\langle n_e\rangle)$;  Yellow solid lines: the convective momentum transport and magnetic momentum transport $m_e\langle \mathbf{U} \cdot \nabla U_{ez}\rangle/e -(\langle \mathbf{U}_{e\perp}\rangle\times \langle\mathbf{B}_{\perp}\rangle )_z/c $. }
\label{ohm_xy}
\end{figure} 
We show  $z$-averaged terms in Eq.~(\ref{eturbz}) at $\Omega_{i0} t=$ 0.05, 0.075, and 0.1 in Fig.~\ref{ohm_xy}. At $\Omega_{i0} t=0.05$ when Buneman instability just starts, the mean electric field $E_z$  is nearly zero within the current sheet.  However, at $\Omega_{i0} t \sim 0.075$ when the instability peaks, the mean electric field significantly deviates from zero, and $\langle E_z \rangle$ is almost completely supported  by inertia $-m_e \partial_t \langle U_{ez}\rangle/e$ and  drag $\langle D_{ez}\rangle$, i.e. $\langle E_z\rangle\approx-m_e \partial_t \langle U_{ez}\rangle/e+ \langle D_{ez}\rangle$.   At $\Omega_{i0} t \sim 0.1$ when turbulence decays, drag and turbulence induced dissipations also become weaker compared to those at the peak of the turbulence developement. The mean electric field is still supported by inertia and drag around the mid-plane $y \sim 0$. Contributions from other terms in the Ohm's Law are all negligible compared to inertia and drag.  Therefore, when the Buneman turbulence is strong, i.e. around peak of the instability, Eq.~(\ref{eturbz}) can be simplified as $\langle E_z\rangle\approx-\dfrac{m_e}{e} \partial_t \langle U_{ez}\rangle+ \langle D_{ez}\rangle$.
\begin{figure}
\includegraphics[scale=0.9, trim=0 50 0 80,clip]{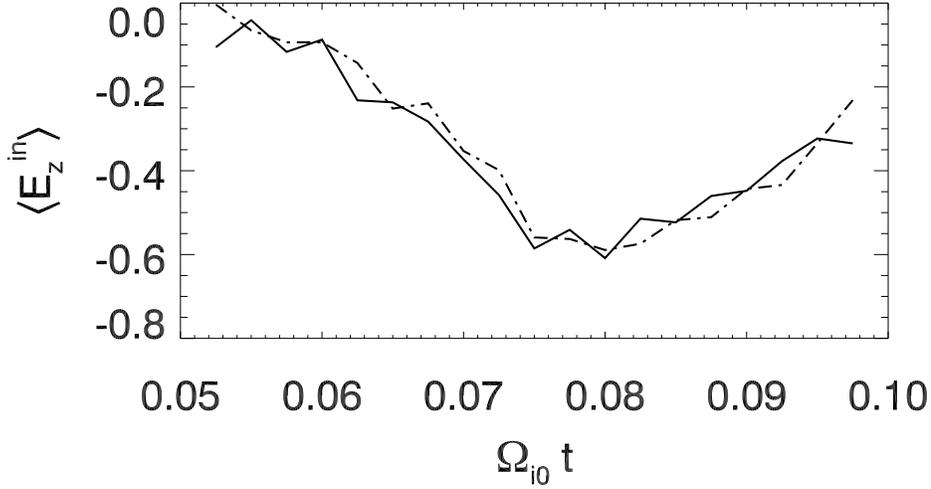} 
\caption{ $\langle E_z^{in}\rangle$ (solid line) is calculated from the mean magnetic flux $\langle A_z\rangle$  using Coulomb gauge i.e. $\langle E_z\rangle=-\partial_t \langle A_z\rangle/c$ while the mean electric field $\langle E_z\rangle$ shown as dashed line is extracted directly from the simulation.}
\label{induct}
\end{figure} 

This mean electric field is an important consequence of turbulent dissipation.  Usually we focus on the dissipation of kinetic energy of electron streams, and ignore the fact that the magnetic field associated with the electron streams also decays since it is determined by the current density $j_z =j_{ez}+j_{iz} \sim  j_{ez}$ and $(\nabla\times \mathbf{B})_z = 4\pi j_{ez}/c$, here we neglect the contribution from the time variation of the electric field that is much weaker compared to $j_{ez}$. The decay of the magnetic field induces an electric field $ E_z^{in}=-\partial_t A_z/c$ (Coulomb gauge). Indeed, as shown in Fig.\ref{induct}, the mean inductive electric field $\langle E_z^{in}\rangle$ calculated from the magnetic flux $A_z$ obtained from the simulations matches very well with  $\langle E_z\rangle$ observed in the simulation. As a result, we have 
\begin{equation}
\langle E_{z}^{in} \rangle = -\dfrac{m_e}{e}\partial_t \langle U_{ez}\rangle+ \langle D_{ez} \rangle.
\label{ein}
\end{equation}
Drag  generated by Buneman instability not only dissipates the kinetic energy of electron beams but also the associated magnetic energy that induces the electric field. 

We can show with our simulation that when the instability saturates the non-spatial averaged inductive electric field $E_z^{in}$ itself also equals to the sum of inertia and drag: 
\begin{equation}
E_{z}^{in}  = -\dfrac{m_e}{e}\partial_t  U_{ez} +  D_{ez}.
\label{einnav}
\end{equation}

\subsection{Local Quasi-static effects:  Anomalous Momentum Transports and Buneman Waves}
\label{sxz}
 We now study the local effects and look at the z-component of Eq.~(\ref{turbflu}) in the mid-plane of the current sheet:
\begin{equation}
%\begin{multline}
  E_z =  D_{ez} -\dfrac{m_e}{e}(\partial_t  U_{ez}  + U_{ez} \partial_z U_{ez}) -\dfrac{1}{e\langle n_e\rangle} \partial_z P_{ezz}.
\label{bunohm}
%\end{multline}
\end{equation}
In this equation we have used $(\mathbf{U}_e\times \mathbf{B})_z=0$  in the mid-plane of the current layer, and the contribution from non-diagonal pressure is negligible. Using Eq.~(\ref{einnav}) we can rewrite the equation as
\begin{equation}
E_z = E_z^{in} + E_z^{wv},
\label{esplit}
\end{equation}
where
\begin{equation}
  E_z^{wv} = 
  -\dfrac{m_e}{e} U_{ez} \partial_z U_{ez} -
   \dfrac{1}{e\langle n_e\rangle} \partial_z P_{ezz}.
\label{2ohm}
\end{equation}
$E_z^{wv}$ is the localized electric field generated by Buneman instability, and satisfies $\langle E_z^{wv}\rangle=0$. 

\begin{figure}
\includegraphics[scale=1, trim=100 30 0 0,clip]{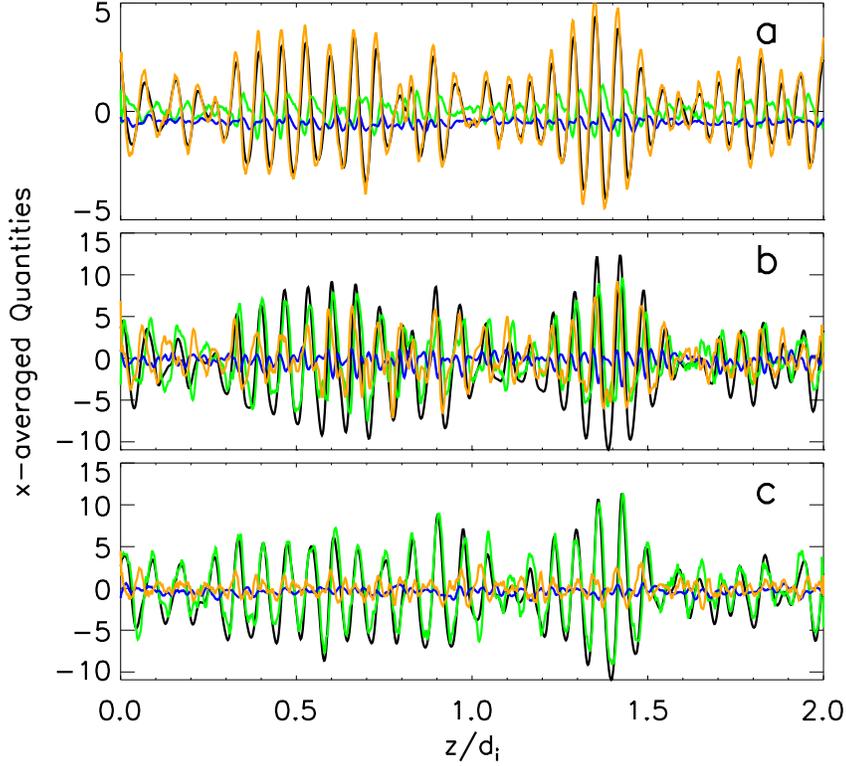} 
\caption{Terms in the Ohm's law as a function of $z$ at $\Omega_{i0} t=$ 0.05 (panel a), 0.075 (panel b), and 0.1 (panel c) in the mid-plane of the current layer. Black lines: the total electric field $E_z$;  Orange lines: the electron convective momentum transport $m_e U_{ez} \partial_z U_{ez}/e$;  Green lines: the thermal momentum transport $ -\partial_z P_{ezz}/(e \langle n_e\rangle)$;  Blue lines: the RHS of Eq.~(\ref{einnav}) . 
   }
\label{ohmxz}
\end{figure} 

In Fig.~\ref{ohmxz} we show each of the terms in Eq.~(\ref{bunohm}), i.e., the convective momentum  transport $m_e U_{ez} \partial_z  U_{ez}/e$,  the thermal momentum transport $ -\partial_z  P_{ezz}/(e \langle n_e\rangle) $, and $E_z$ as a function of $z$ at $\Omega_{i0} t=$ 0.05, 0.075 and 0.1. We also show the RHS of Eq.~(\ref{einnav}) which equals to $E_z^{in}$. We examine their relative contributions to balance the total electric field $E_z$. 
At all times the turbulence in $z$-direction is dominated by the fastest growing waves of Buneman instability. Because of the very low phase speed of the Buneman waves, the shapes of waves do not appear to vary significantly, only amplitudes of waves change.  
   
 At $\Omega_{i0} t=0.05$, the convective momentum transport contributes most to the total electric field $E_z$, while the contribution from thermal momentum transport $- \partial_z P_{ezz}/(e \langle n_e\rangle)$ is small. Initially the velocity is uniform along $z$, thus the strong convective momentum transport is caused by the Buneman instability that feeds the growth of waves. At this time, the Buneman instability is still at its linear stage and waves only absorb the energy of resonant electrons. Electron trapping is weak and thus heating is weak too. 
 
 At $\Omega_{i0} t=0.075$, near the saturation of Buneman instability, while the convective momentum transport remain roughly the same, the thermal momentum transport increases by more than a factor of 10 compared to that at $\Omega_{i0} t =0.05$. This results in a significant increase of the amplitude of the total electric field $E_z$. Given that the initial electron pressure is uniform and isotropic, the thermal momentum transport is driven by Buneman instability (anomalous thermal momentum transport). This implies that the energy conversion from electron streaming energy to thermal energy is strong. 
 
At $\Omega_{i0} t=0.1$, the Buneman instability decays and the anomalous thermal momentum transport almost fully supports the Buneman waves while the electron convective momentum transport decreases to near zero. With the decay of Buneman instability, the anomalous thermal momentum transport decreases with the Buneman waves. In Paper I, we have shown that at $\Omega_{i0} t =0.075$ to 0.1, fast adiabatic phase mixing takes place. The non-adiabatic and irreversible trapping and de-trapping transfer the energy of electrons gained from waves into electron heat. Therefore, it's not surprising that the anomalous thermal momentum transport rapidly takes over the electron convective momentum transport. 

Note that the amplitude of the RHS of Eq.~(\ref{einnav}) (blue line)  is in general much smaller than that of $E_z$ and has an negative sign on average. This means that $E_z^{in}$ accelerates electrons on average. The total of the RHS of  Eq.~(\ref{einnav}) and  Eq.~(\ref{2ohm})  matches $E_z$ as expected (not shown in Fig.\ref{ohmxz}).

Eq.~(\ref{2ohm}) determines the growth of the Buneman waves.  We can explicitly approximate the electron velocity as $U_{ez}\approx\langle U_{ez}\rangle \pm \sqrt{e\phi^{wv}/m_e}$, where  $E_{z}^{wv}=-\partial_z \phi^{wv}$, and the first term in Eq.~(\ref{2ohm}) becomes $E_z^{wv}/2\pm \langle U_{ez}\rangle \partial_z \sqrt{m_e\phi^{wv}/e}$. This implies that the convective momentum transport not only supports the waves by trapping electrons but also supplies the thermal momentum transport with transferring de-trapped electrons. Therefore the growth of waves stops  when the thermal momentum transport takes over the convective momentum transport, i.e. $\vert m_e U_{ez}\partial_z U_{ez}\vert<\vert\partial_z P_{ezz}/\langle n_e\rangle\vert$ that implies $m_e U_{ez}^2/2<P_{ezz}/\langle n_e \rangle$. In fluid theory, the growth of Buneman instability can not stop due to the lack of dissipation generated by wave-particle interactions.

In our simulation $v_{te}^2=T_{ezz}/m_e$, and $P_{ezz}/\langle n_e \rangle\sim T_{ezz}$.  The criteria for saturation is $U_{ez} < 2 v_{te}$,  which is the same as the threshold to trigger Buneman instability in linear kinetic theory $U_{ez} > 2 v_{te}$. \cite{papa77rgsp} 

Integrating Eq.~(\ref{2ohm}) over $z$, we have:
\begin{equation}
 \dfrac{m_e}{2e} U_{ez}^2 + \dfrac{1}{e\langle n_e\rangle} P_{ezz}- \phi^{wv}=C(t),
 \label{dber}
 \end{equation}
 where $\langle \phi^{wv}\rangle=0$ and $C(t)$ is a function of time.  Eq.~(\ref{dber}) is a Bernoulli-like equation, implying that Buneman instability is locally quasi-static. This is consistent with the basic feature of adiabatic phase mixing of electrons near the saturation of Buneman instability: the growth rate of the Buneman waves is much slower than the bounce rate of trapped electrons.
  
\subsection{The Coupling between Micro-Macro Processes}
Eq.~(\ref{einnav}) and (\ref{2ohm}) are two separable processes that describe the global dissipation and localized momentum transports respectively. We now show the importance of drag in linking the localized momentum transport and the global energy dissipation.

Multiplying $ n_e /\langle n_e\rangle$  to both sides of Eq.~(\ref{2ohm}) and average along $z$-direction, we obtain:
\begin{eqnarray}
\langle D_{ez} \rangle = 
  \dfrac{m_e}{e} \langle  n_e \partial_z U_{ez}^2/2\rangle +
   \dfrac{1}{e\langle n_e\rangle^2}\langle  n_e \partial_z P_{ezz}\rangle. 
   \label{DT}
\end{eqnarray}
where we have applied $\langle n_e E_z^{wv}\rangle/\langle n_e\rangle=-\langle D_{ez}\rangle$.

Eq.~(\ref{DT}) shows that the drag is the origin of global momentum transports. We have shown in Eq.~(\ref{2ohm}) and Fig.~\ref{ohmxz} that the local convective momentum transport feeds the growth of the Buneman waves by trapping and the trapping quickly converts the absorbed kinetic energy into thermal energy. Thus the local thermal momentum transport plays a competitive role against the local convective momentum transport. As a result, the global electron convective momentum transport is weak while the global thermal momentum transport dominates because the de-trapped electrons are free to bring the local thermal momentum away from where it is generated.

\begin{figure}
\includegraphics[scale=1, trim=100 110 0 50,clip]{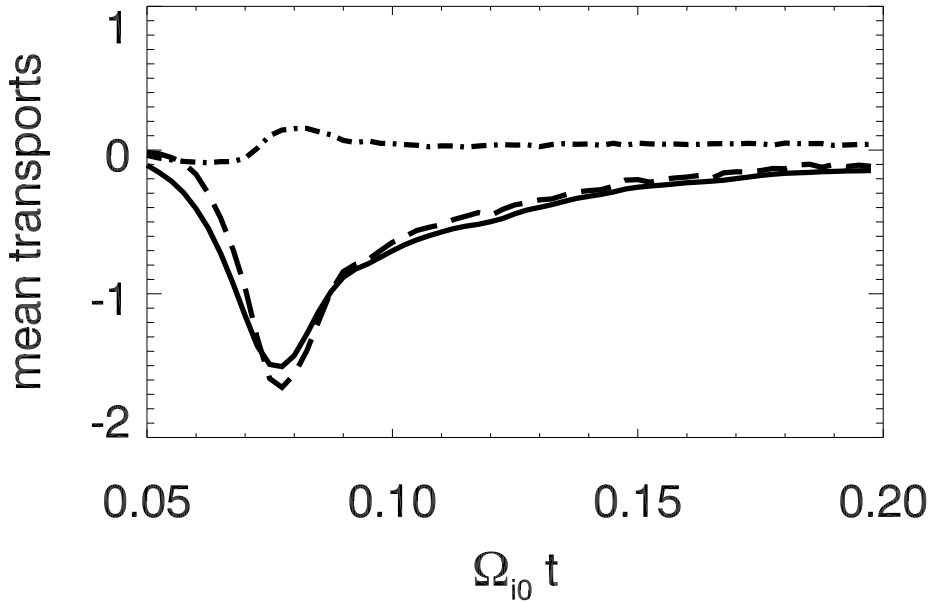} 
\caption{Solid line is the mean drag $\langle D_{ez}\rangle$, the dashed line is the mean thermal momentum transport and the dot-dashed line is the mean convective momentum transport. }
\label{fDT}
\end{figure}

Each term in Eq.~(\ref{DT}) calculated from our simulation is shown  in Fig.~\ref{fDT}. As we expect, the mean drag is nearly balanced by the mean thermal momentum transport while the mean convective momentum transport is much smaller than the thermal momentum transport. The drag links the adiabatic thermalization of electrons inside the solitary waves to the global irreversible heating process. 

\section{ Thermalization of Kinetic Energy }
\label{thermal}
In this section, we establish a closure for pressure by using energy conservation in the mid-plane. Along with  Eq.~(\ref{einnav}), (\ref{2ohm}) and continuity equation, we have a full EMHD description of the ``1D" Buneman instability.

\begin{figure}
\includegraphics[scale=0.9,trim=30 80 0 0,clip]{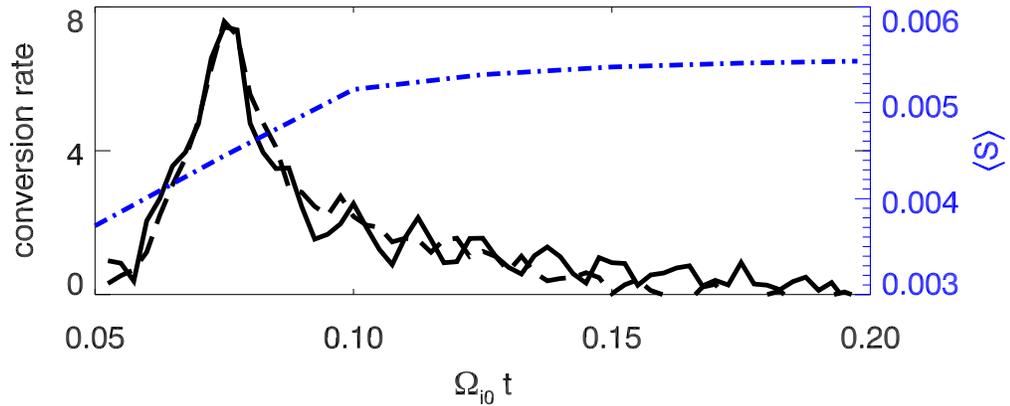}
\caption{ The black solid line represents the changing rate of kinetic energy of electrons $-m_e\partial_t \langle n_e U_{ez}^2/2\rangle $ (so plotted to allow easy comparison with the temperature increase) and the black dashed line represents the temperature changing rate $\partial_t \langle P_{ezz}\rangle/2$ . The blue dot-dashed line represents the average Boltzmann entropy $\langle S\rangle$. }
\label{fconversion}
\end{figure}

 The mean energy density in a 2D current layer as a function of $y$ is $\overline{W}(y) = (\int_A B^2/(8\pi) dxdz + \int_A E^2/(8\pi) dxdz+\sum_0^{N} m_e v_{e}^2/2)/A=\langle B^2/(8\pi)\rangle + \langle E^2/(8\pi)\rangle + \langle m_e n_e U_e^2/2\rangle +\langle (P_{exx}+P_{eyy}+P_{ezz})/2\rangle$, where  $v_{e}$ is the velocity of each electron, $A$ is the simulation area in $xz$-plane, and $N$ is the total electron number. We have neglected ion contributions. In the mid-plane Buneman instability does not explicitly involve magnetic field because $(\mathbf{U}_e \times \mathbf{B})_z=0$.  We compare the remaining terms of the mean energy density in Fig.~\ref{fconversion}. It is clear that at all times the decrease of electron kinetic energy is balanced by the increase of thermal energy, while the electric energy remains negligible (see Fig. 1), i.e. 

 \begin{eqnarray}
\partial_t (\langle m_e n_e U_{ez}^2 \rangle + \langle P_{ezz}\rangle) = 0. 
\label{conversion}
\end{eqnarray}
Eq.~(\ref{conversion}) is locally approximately valid, i.e. $\partial_t (m_e n_e U_{ez}^2  + P_{ezz}) \approx 0 $, or the energy density roughly conserves locally. This is because the energy exchanges between electrons and waves occur in highly localized solitary waves and the exchanges are very efficient. This equation together with  Eq.~(\ref{einnav}) and Eq.~(\ref{2ohm}) provide a set of fluid description for the macroscopic effects produced by wave-particle interactions in 1D Buneman instability.

We have shown in \S~\ref{sxz} that the criteria for Buneman instability to saturate is $m_e  U_{ez}^2/2\leqslant P_{ezz}/\langle n_e\rangle$. From Eq .~(\ref{conversion}), we have $m_e n_0U_{ez,0}^2-m_en_eU_{ez}^2=P_{ezz}-P_{ezz,0}$, and using  $m_e U_{ez}^2/2=P_{ezz}/\langle n_e\rangle$ at saturation, we find  $U_{ez}\sim 7 v_A$ and $P_{ezz}=0.28$ at $\Omega_{i0} t=0.075$, the time when the instability saturates.  These agree with the simulation results shown in Fig.~\ref{diss2}, where the initial drift $U_{ez,0}=9 v_A$, $P_{ezz}=0.04 m_i v_A^2$ and $\langle n_e\rangle = n_0 $ where $n_0$ is the background density. 

 As we have shown in Paper I, the conversion of kinetic energy to thermal energy due to trapping and de-trapping is irreversible. This can be seen in the monotonic increase of the average Boltzmann entropy $\langle S\rangle=-\int_0^{L_z} f(v_z, z) lnf(v_z, z) dv_z dz/L_z$, where $f$ is the electron distribution function, also plotted in Fig.~\ref{fconversion}. The entropy shows a significant increase $\sim 38\%$ during $\Omega_{i0} t=0.05-0.1$. 

\section{Dissipation of Magnetic Energy In the Thin Current Sheet}
Dissipation of kinetic energy must be accompanied by the loss of magnetic energy associated with the current. According to the Ampere's law the magnetic energy is $ B_x^2/(4\pi)^2\sim j_{ez}^2/\omega_{pe}^2$, where we used $\delta y\sim d_e$, and $d_e$ is  the width of the current sheet. The magnetic energy loss is therefore $\Delta B_x^2/(8\pi)\sim m_e \Delta n_e U_{ez}^2/2$.  The Ampere's law also implies that the damping of  magnetic energy $B_x^2/(8\pi)$ occurs in layers away from $y=0$. Thus in the mid-plane,  while the inductive electric field $E_z^{in}$ due to the decay of magnetic field is important, magnetic energy decay cannot be studied only within the mid-plane.  So far we have been focusing only on the $z$-component equations in the mid-plane because in this plane Buneman waves propagate primarily in $z$-direction. This property of Buneman waves greatly simplifies the problem and allow us to treat it justifiably as in ``1D".  To account for the dissipation of magnetic energy, however, we have to examine the $x$ and $y$-components of fields and thermal pressure produced by heating. 
%A question naturally arises is how good is our 1D approximation to Buneman instability in and away from the mid-plane when the dissipation of magnetic energy is not 1D? }

Above or below the mid-plane, velocity shear along $y$ can cause Buneman instability to become slightly oblique in the $yz$-plane. \cite{che11pop} In force-free current sheet,  $j_{ex}/j_{ez}=-B_x/B_z$,  thus the electron drift becomes more and more oblique as $y$ increases. Therefore, Buneman wave away from the mid-plane has all three electric field components as it propagates along the magnetic field.\cite{che10grl}  As a result electron heating is in directions both parallel and perpendicular to the magnetic field. In the following we discuss the relation between the magnetic energy damping and the electron heating.
%and whether 1D approximation in $z$ is valid in the mid-plane and the planes away from $y=0$. }
 \begin{figure}
\includegraphics[scale=0.8,trim=50 70 0 0,clip]{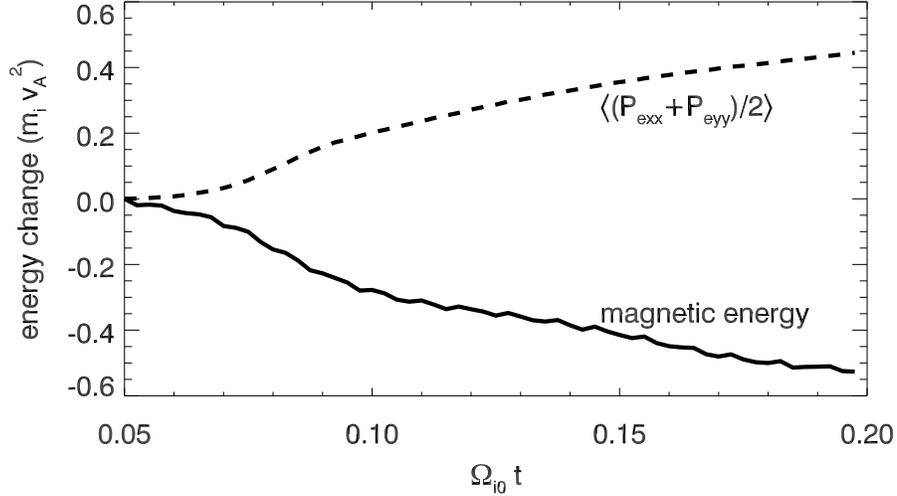}
\caption{The solid line is magnetic energy $ \overline{W}_{B}(y=0, t)-\overline{W}_{B}(y=0,t=0)$ and the dashed line is $ \overline{W}_{P_{x,y}}(y=0, t)-\overline{W}_{P_{x,y}}(y=0,t=0)$. }
\label{B_y0}
\end{figure}

Within the thin current sheet, $\vert j_{ex}/j_{ez}\vert << 1$, $\vert j_{ey}/j_{ez}\vert << 1$ and $\vert B_y/B_0\vert<<1$, thus $x$ and $y$ components of the inertia term in electron momentum equation Eq.~(\ref{2flu}) and $B_y$ are negligible.  The $x$ and $y$ components of  Eq.~(\ref{2flu}) become:
\begin{equation}
\begin{array}{lr}
 n_e e E_x  = \partial_x(B_x^2+B_z^2)/(8\pi) -\partial_x P_{exx} , \\
 n_e e E_y  = \partial_y(B_x^2+B_z^2)/(8\pi) -\partial_y P_{eyy}.
 \end{array}
 \label{enxy}
\end{equation}
The inhomogeneity of magnetic field and $P_{exx}$ is due to the increase of $B_x/B_z$ with $y$. The propagation of Buneman waves deviate from $z$ in the $xz$ plane with the increase of $y$. Eq.~(\ref{enxy}) tells us that the perpendicular electric fields convert magnetic energy into thermal energy to produce perpendicular thermal pressure, i.e. the dissipated magnetic energy produces perpendicular heating in the thin current, or  $\overline{W}_{B}+\overline{W}_{P_{x,y}}= \overline{W}_B(t=0)+\overline{W}_{P_{x,y}}(t=0)$ where $\overline{W}_B=\langle (B_x^2 + B_z^2)\rangle /8\pi$ and $\overline{W}_{P_{x,y}}=\langle (P_{exx}+P_{eyy})/2\rangle$. As y get close to $2w_0$, the edge of the current sheet, electric fields become very weak. Consequently the heating is very weak and the magnetic pressure $B^2/8\pi$ is approximately constant in time.

 In the mid-plane $B_x = 0$ and the loss of magnetic energy $B_z^2$ is balanced by heating in $x$ and $y$.  The loss of average magnetic energy $ \Delta \overline{W}_{B}= \Delta \langle B_z^2/8\pi \rangle$ is countered by the gain of $\Delta \overline{W}_{P_{x,y}}=\Delta \langle (P_{exx}+P_{eyy})/2\rangle$, as shown in Fig.~\ref{B_y0}.  
\begin{figure}
\includegraphics[scale=0.8,trim=0 70 30 50,clip]{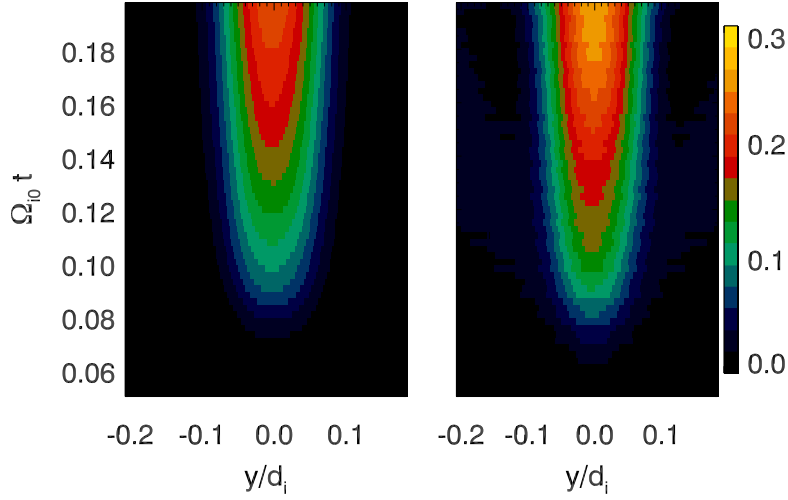}
\caption{The left panel is $ \overline{W}_{B}(t)- \overline{W}_{B}(t=0)$ and the right panel is $  \overline{W}_{P_{x,y}}(t)- \overline{W}_{P_{x,y}}( t=0)$. }
\label{B_y}
\end{figure}
We further show the time evolution of average magnetic energy loss $ \overline{W}_{B} (y,t)-\overline{W}_{B}(y,t=0)$ and the average thermal energy gain $\overline{W}_{P_{x,y}}(y,t)-\overline{W}_{P_{x,y}}(y, t=0)$  along $y$  in Fig.~\ref {B_y}. To allow easy comparison, the absolute values of average magnetic energy loss is shown. It is obvious that the two agree with each other in the thin current layer. At the edge of the current $y \sim 2 w_0\sim 0.2d_i$, the loss of magnetic energy and heating become nearly zero. Therefore, we have,  
\begin{gather}
\Delta \langle P_{exx} \rangle + \Delta \langle P_{eyy}\rangle \sim \Delta \langle P_{ezz}\rangle,\\
\Delta \langle B_z^2/8\pi\rangle + \Delta \langle B_x^2/8\pi \rangle\sim -(\Delta \langle P_{exx}\rangle +\Delta \langle  P_{eyy}\rangle)/2,\\
B_z^2+B_x^2 \approx constant, \; for\: y > 2w_0.
\end{gather} 

The change of $P_{ezz}$ equals to the loss of the electron kinetic energy $\Delta P_{ezz} \sim \Delta (m_e n_e U_{ez}^2) $.  The latter leads to the loss of magnetic energy via the Ampere's Law $\Delta U_{ez} \sim -\Delta j_{ez}/(en_e)\sim -c\Delta B/(4\pi w_0 en_e)$. Therefore, 
\begin{equation}
\Delta P_{ezz}/\Delta P_{\perp}\sim \Delta (m_e n_e U_{ez}^2)/\Delta B^2/8\pi\sim (d_e/w_0)^2.
\label{equpar}
\end{equation}
Thus the equipartition between parallel and perpendicular thermal energy is a direct consequence of the Ampere's Law when the current sheet is of electron scale, i.e $w_0\sim d_e$.

In principle, at each layer with $y=y_0$, we can apply our 1D $z$-component fluid description of Buneman turbulence and the corresponding parallel heating the same way as we do at $y=0$. However, the time evolution of the magnetic field and heating  is beyond the scope of this paper since it requires a full 3D model of  Buneman instability. 
%------------
\section{The Influence of Ions }
\label{ioninfluence}
%------------
 We have found the electron fluid description (Eq.~(\ref{einnav}), (\ref{2ohm}) and (\ref{conversion})) of the macroscopic effects produced by the wave-particle interactions in Buenman instability. We have so far neglected the dynamics of ions for the following reason: The time scale of Buneman instability $\sim (m_i/m_e)^{1/3} \omega_{pe}^{-1}\sim (m_e/m_i)^{1/6} \omega_{pi}^{-1} $ is much smaller than the ion gyro-period $\Omega_{ci}^{-1}= c\omega_{pi}^{-1}/v_{A}$  and similar to the ion dynamical responding time scale $\sim \omega_{pi}^{-1}$.  On the other hand the time scale of the instability is comparable to the electron gyro-period $\Omega_{ce}^{-1}= \omega_{pi}^{-1}m_e c/(m_i v_{A}) $ and much longer than the electron dynamical responding time scale $\omega_{pe}^{-1}=(m_e/m_i)^{1/2} \omega_{pi}^{-1}$. Thus energy exchanges primarily between waves and electrons rather than with ions, the thermalization generated by trapping and de-trapping of ions is much weaker than that of electrons and the wave energy loss to ions  can be neglected--- this is consistent with the approximate conservation of the total energy in electron fluid description during Buneman instability. 
 
 During the Buneman instability the oscillation of ions in waves facilitates the energy exchange between waves and electrons but the heating of ions is negligible. Therefore the ion momentum equation can be simplified as 
\begin{equation}
E_z=\dfrac{m_i}{e}\partial_t  U_{iz}+ D_{iz}.
\label{ion}
\end{equation}
 
It should be noted that Buneman instability is triggered by the relative drift between electrons and ions. 
%If the ion's drift is very small compared to electron's drift, we can approximate it is in the ion's rest frame. 
In the case that the ions' drift is non-zero, we must replace $U_{ez}$ by $U_{ez}-U_{iz}$ where $U_{iz}$ is the ion drift, and pressure by $P_{ezz}+m_e n_e U_{iz}^2 -2m_en_e U_{ez}U_{iz}$ in the electron fluid equations we obtained. The ion drift does not affect the dissipation of magnetic energy since the current sheet is determined by the relative drift $U_{ez}-U_{iz}$. 
  
To summarize, we list  the compete set of equations in ion rest-frame that can naturally trigger Buenman instability and determine its evolution:
\begin{gather*}
\label{buneq1}
E_z=\dfrac{m_i}{e}\partial_t  U_{iz}+ D_{iz},\\
E_z=-\dfrac{m_e}{e}(\partial_t  U_{ez}+ U_{ez}\partial_z U_{ez}) -\dfrac{\partial_z P_{ezz}}{e n_e}+D_{ez},\\
 E_z^{wv}  = -\dfrac{m_e}{e}U_{ez} \partial_z U_{ez}
-\dfrac{1}{e\langle n_e\rangle} \partial_z P_{ezz}, \\
\partial_t n_{s} + \partial_z (n_s U_{sz})=0,\\
\partial_z E_z^{wv}=4\pi\sum_s q_s n_s,\\
 \partial_t (m_e n_e U_{ez}^2 + P_{ezz})=0,\\
 D_{sz}=-\delta n_s E_z^{wv}/n_0,\\
 n_s=n_0+ \delta n_s,
\label{buneq}
\end{gather*}
where we use gauge $\nabla\cdot \mathbf{A}=0$ .  Eq.~(\ref{bunohm}) replace Eq.~(\ref{einnav}) so that both electron and ion use momentum equations.
 \section{Summary and Discussions }
\label{diss}
In this paper we have studied the macroscopic momentum transports and energy dissipation generated by wave-particle interactions in Buneman instability in the mid-plane of a thin current layer with a guide magnetic field. This study is important for the understanding of the role of diffusion region kinetic turbulence in magnetic reconnection. Using PIC simulations and detailed analysis of electron fluid equations, we found 
\begin{enumerate}
\item Buneman electrostatic waves propagate along the magnetic field and leads to parallel momentum transports and dissipation of electron kinetic energy.  In the mid-plane, Buneman instability behaves like a 1D problem along the guide field $B_z$. 
\item The global energy dissipation and local momentum transports during Buneman instability are two separable processes and the electric field generated by Buneman instability can be separated into two components: the low frequency inductive electric fields $E_z^{in}$ and high frequency turbulence fluctuations $E_z^{wv}$.  As a result, the electron momentum equation (Eq.~\ref{turbflu}) that incorporates turbulence drag  is split into two equations for $E_z^{in}$ and $E_z^{wv}$ respectively. The first equation (Eq.~\ref{einnav}) describes the global damping of electron kinetic energy produced by drag and the acceleration of electrons produced by $E_z^{in}$. $E_z^{in}$ is induced by the loss of the magnetic energy associated with the electron streams. The second equation (Eq.~\ref{2ohm}) describes the macroscopic balance in the localized Buneman solitary waves among the electric force, the local convective momentum and thermal momentum transports. A different form of Eq.~(\ref{2ohm}), i.e. Eq.~(\ref{dber}), is similar to the well known Bernoulli equation in fluid mechanics, a direct consequence of the locally quasi-static nature of Buneman instability. This equation can stop the growth of Buneman instability. The Buneman instability saturates when the drift decreases below the threshold of Buneman instability.
\item Drag couples local momentum transports with global energy dissipation, and links the microscopic heating process inside the localized Buneman solitary waves to the macroscopic kinetic energy dissipation of electrons.
\item The dissipated kinetic energy of electron stream is converted into the parallel electron heat along the magnetic field in the mid-plane. The local conservation of total energy is a result of the very efficient energy exchanges between electrons and solitary waves during Buneman instability.  This condition truncates the infinite moments of fluid equations. Thus, we have found a set of equations, including Eq.~(\ref{einnav}), (\ref{2ohm}) and (\ref{conversion}), for the macroscopic effects of Buneman instability in the mid-plane of a thin current layer. The electron fluid equations together with cold ion equations form a complete description of Buneman instability as listed in \S~\ref{ioninfluence}.
\item If  the drift of ions $U_{iz}$ is non-zero, the electron fluid equations for Buneman instability should be transferred to the ion's rest frame by replacing $U_{ez}$ and pressure $P_{ezz}$ by  $U_{ez}-U_{iz}$ and $P_{ezz}+m_e n_e U_{iz}^2 -2m_e n_e U_{ez}U_{iz}$ respectively. 
\item Dissipation by Buneman turbulence is irreversible as seen in the monotonic increase of Boltzmann entropy. The fastest increase of entropy occurs at the time when the growth of Buneman instability  peaks.  
\item Magnetic energy dissipation is associated with the perpendicular components of Buneman waves. The magnetic energy is converted into electron thermal energy as shown in the increases of the perpendicular components of the pressure tensor. The process is decoupled from the parallel  heating. The ratio of perpendicular and parallel heating rate is proportional to $(d_e/w_0)^2$. The observed equipartition of heating rate between parallel and perpendicular directions in our simulation is a result of the width of the current layer being $\sim d_e$.  
\end{enumerate}

It is useful to highlight the similarities and differences between joule heating produced by collisions and turbulence heating caused by wave-particle interactions -- or drag as it's macroscopic manifestation.  Both drag and collisions can dissipate kinetic energy and cause the increase of the temperature and entropy, but the underlying physics are different: 1) Drag is generated by wave-particle interactions while collision is generated by particle-particle interactions; 2) Drag is the feature of kinetic instabilities that produces non-equilibrium structures, such as localized intense electric field and non-Maxwellian velocity distribution, while collisions tend to drive the system to equilibrium and produce Maxwellian velocity distribution; 3) Heating induced by drag has a time lag $\tau_{bun}$ in the conversion of convective momentum to thermal momentum during the growth of Buneman waves. The time lag is of the Buneman turbulence time scale $\tau_{bun} \sim 1/\omega_{pe}$.  Compared with collisions, $\tau_{bun}$ is much shorter than the collision time scale $\tau_c >>1/\omega_{pe}$.

The effects of turbulence dissipation is commonly parameterized as effective anomalous resistivity $\eta_{eff}$ in MHD theory. In this parameterization drag assumes a resistivity-like form $D_{ez}= \eta_{eff}  j_{ez}/ n_e$,  and the dissipation rate has the simplest form of joule heating, i.e., $\partial_t P_{ezz}\sim\eta_{eff} j_{ez}^2/n_e$.  We can see that in this parameterization $\partial_t P_{ezz}$ depends on kinetic energy density rather than the changing rate of kinetic energy density as we have found for Buneman instability. As a method to estimate the level of anomalous heating if we do not know the underlying physics, parameterization is still the simplest and most effective method.

In most cases, we are only interested in the heating effect of Buneman instability rather than the form of  Buneman waves. In such cases only  Eq.~(\ref{einnav}) and (\ref{conversion}) associated with the global effects are useful, but we must give $D_{ez}=f(t)$, which can be obtained either with kinetic theory or fitting of PIC simulations. Given $D_{ez}$ we have
 \begin{flalign} 
U_{ez}=\int f dt+U_{ez0},\\
P_{ezz}=n_0U_{ez0}^2-n_e U_{ez}^2,
 \end{flalign} 
where $U_{ez0}$ is the initial drift of electron beams.

The ultimate question is whether turbulence dissipation/heating can accelerate magnetic reconnection.  Comparing with the time scale of large scale magnetic reconnection $ \tau_{reconn} >>d_i/v_{A0} \sim 1/\Omega_{i0} $, $\tau_{bun}$ is still quite short. This implies that  anomalous heating on kinetic scale has the potential to impact on large scale reconnection. This point will be addressed in a future paper.

\begin{acknowledgements}
The author like to thank the anonymous referees whose comments have helped in many improvements of this manuscript. The author gratefully thanks inspiring discussions with P. H. Diamond and M. V. Goldman and helpful comments from M. Swisdak on the manuscript. The author thanks the colleagues in NASA/GSFC  and PPPL for their questions and comments. Finally the author is very grateful to M. L. Goldstein for his careful reading of this manuscript.

This research is supported by the NASA Postdoctoral Program at NASA/GSFC administered by Oak Ridge Associated Universities through a contract with NASA and NASA grant NNH11ZDA001N. The simulations and analysis were partially carried out at the National Energy Research Scientific  Computing Center and at NASA/Ames High-End Computing Capacity. 
\end{acknowledgements}

\appendix

\end{document}